# A COMPARISON STUDY OF OBJECT-ORIENTED DATABASE CLUSTERING TECHNIQUES

To appear in *Information Sciences*

**Jérôme Darmont**
Blaise Pascal University
Institut des Sciences de l'Ingénieur
Clermont-Ferrand, France
darmont@libd1.univ-bpclermont.fr

**Le Gruenwald**
University of Oklahoma
School of Computer Science
Norman, Oklahoma 73019
gruenwal@mailhost.ecn.uoknor.edu

***Abstract:*** It is widely acknowledged that a good object clustering is critical to the performance of OODBs. Clustering means storing related objects close together on secondary storage so that when one object is accessed from disk, all its related objects are also brought into memory. Then access to these related objects is a main memory access that is much faster than a disk access. The aim of this paper is to compare the performance of three clustering algorithms: Cactis, CK and ORION. Simulation experiments we performed showed that the Cactis algorithm is better than the ORION algorithm and that the CK algorithm totally outperforms both other algorithms in terms of response time and clustering overhead.

<u>Keywords</u>: Cactis, CK, Clustering, Object-Oriented Databases, ORION, Simulation

## 1. INTRODUCTION

There are several ways to improve response time (i.e., to limit the number of disk Input/Output) in a DBMS. Indexing, clustering (i.e., storing related entities close together on secondary storage) and buffering (i.e., fetching clustered entities at the same time and setting up replacement strategies) are widely used techniques in conventional DBMSs. However, OODBs present additional semantics like structural properties (inheritance, composite objects) and interrelationships between objects. Hence, the existing clustering algorithms (used in relational databases, for instance) have to be adapted to the object-oriented model.

We have chosen to study three clustering algorithms found in the literature: the Cactis [HUDS89], the CK [CHAN89b, CHAN90] and the ORION [BANE87, KIM90a] clustering algorithms. The Cactis and ORION clustering algorithms are already implemented in DBMSs. We have chosen these particular algorithms because we consider them to be different enough from each other to be representative of the current research on clustering techniques in OODBs.

They have also been selected because they present characteristics that are interesting to compare. For instance, CK and ORION are dynamic clustering algorithms as the Cactis clustering algorithm is static. ORION uses only users' hints to cluster a database as the Cactis clustering algorithm uses only statistics about the database and the CK algorithm makes use of both.

Furthermore, the aim of previous performance evaluations performed on these algorithms was only to compare the effects of one particular clustering strategy to those of a "no clustering" policy [CHAN89a, HUDS89]. We intend to compare each of these three algorithms to each other to determine which one performs the best in a given environment. The characteristics that make this algorithm the best should be isolated.

This paper is organized as follows. Section 2 explains the principles of clustering in OODBs. The three studied clustering algorithms are described in Section 3. Section 4 describes our simulation model. In Section 5, the simulation results are analyzed. Section 6 concludes this paper and provides future research directions.

## 2. CLUSTERING IN OODBs

### 2.1. OODB concepts

The notion of object abstraction was first introduced by object-oriented programming languages. Recently, OODBs have added database functionality to this abstraction as an attempt to increase the modeling power and the applicability of databases [TSAN92b]. The object-oriented programming language object abstraction is an extension of the data structure concept with the following basic characteristics:

- *structure:* consisting of components that can be atomic (i.e., flat attributes like integers, reals, or strings), objects (i.e., other objects) or object identifiers (i.e., "pointers" to other objects); unlike data structures, the object state (i.e., values of the structure components) is neither directly changeable nor visible to the user;
- *behavior:* determined by methods, predefined fragments of code that manipulate and export the object state (unlike conventional languages that allow arbitrary code to manipulate data structures);
- *type:* prescribing the "structure" and the "behavior" of an object through the specification of its components and its methods;
- *identity:* naming and locating an object in a manner independent of its state; identity is typically supported identifying objects by an unique number, the object identifier (OID); OIDs are assigned by the system at object creation time and cannot be reused, changed or synthesized.

Conceptually, objects can be viewed as vertices of a directed and possibly cyclic graph: the *Object Graph*. The directed edges of the graph represent the object to object references and they are labeled by the names of their components (Figure 1).

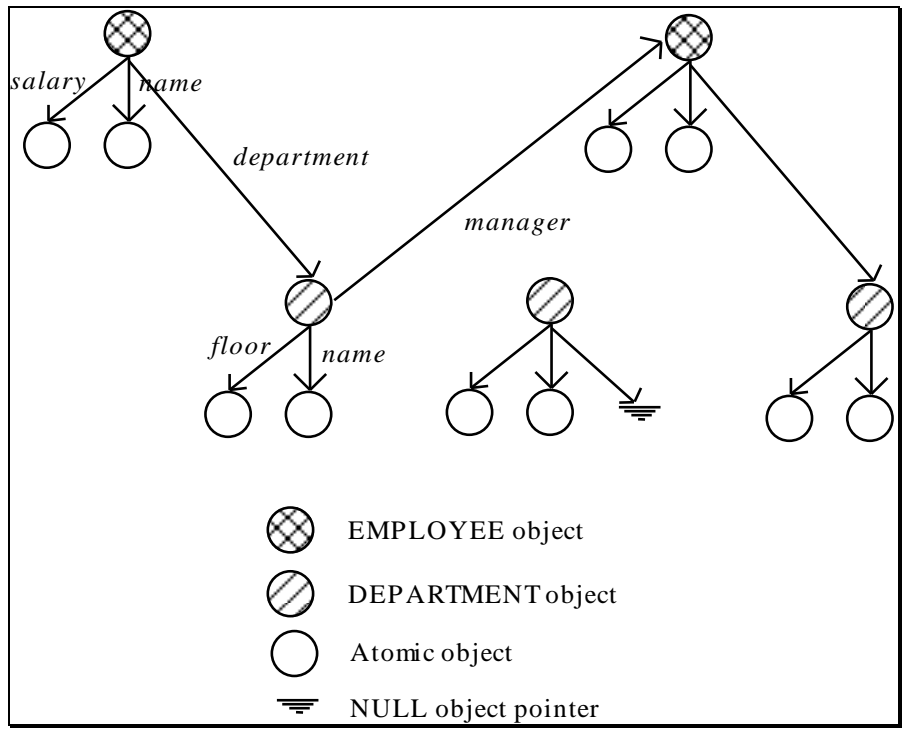

Figure 1: Sample Object Graph

OODBs support additional database functionality:
- *persistence:* i.e., the ability of objects to maintain their state even after the termination of the program that has created them; unlike object-oriented programming languages that only support as many objects as can fit into main memory, OODBs provide access to large collections of objects stored in "stable" secondary storage;
- *storage management:* efficient ways to represent objects in main memory, store them to disk and distribute them to servers; that way OODBs relieve the clients from the burden of storing and retrieving objects from secondary storage, managing main memory as well as the secondary memory;
- *concurrency control and recovery:* to allow concurrent accesses to the objects, and still ensure their integrity; the state of the object base is guaranteed to change only in a consistent manner and is immune to client failures;
- *ad-hoc query facilities:* declarative languages to efficiently apply operations on large sets of objects.

## 2.2. Clustering principles

The goal of object clustering is to reduce the number of disk I/Os for object retrieval. Typically, the unit of data transferred from disk is a page instead of an individual object. If two objects are clustered on the same page, it will take only one disk I/O to access both objects successively [HURS93].

Clustering algorithms attempt to improve the performance of object-oriented database systems by placing on the same page related sets of objects [TSAN92a]. In object-oriented databases, complex objects are the basic units of data manipulation. The subobjects of a complex object may come from different classes. Traditional storage systems tend to group records of the same type physically close to each other on disk. This results in tedious and expensive reconstruction procedures (such as join operations) to retrieve complex objects. Therefore, it is logical to cluster related objects of different classes together to achieve acceptable performance [HURS93].

The problem of clustering can be seen as a graph partitioning problem. The nodes of the graph are the objects and the edges are the links between objects. This problem is NP-complete. However, as the graph of objects represents the database state, all is needed is an incremental solution where new objects are placed at the "right place". Most of the algorithms used can be classified as greedy algorithms: they scan the objects according to their links and try to place them into the same cluster unit [BENZ90].

## 2.3. Clustering strategies

According to [CATT91], clustering in an OODB can actually be performed in many different ways:
- *composite objects:* objects can be clustered according to aggregation relationships;
- *references:* some OODBs allow objects to be clustered according to relationships with other objects; composite objects clustering is, in fact, a special case of this, clustered by aggregation relationships;
- *object types:* objects may also be clustered by their types; if there is a generalization hierarchy, subtype instances may also be clustered in the same segment;
- *indexes:* as in relational DBMSs, it may be possible to cluster objects by an index on their attributes;
- *custom:* some OODBs allow clustering to be performed "on the fly".

Unless objects are stored redundantly, an object can generally be clustered according to one of these rules. Where the rules do not conflict, however, it is possible to follow multiple clustering rules.

Clustering may be performed at two levels:

- *pages:* objects may be clustered according to the smallest physical unit read from disk, which is normally a page; this type of clustering can produce the greatest gains in performance when a "working set" of objects cannot be precisely defined for all applications; page clustering is more useful for clustering by index, reference and composite objects;
- *segments:* objects may be clustered in larger units, when the user is able to specify a meaningful logical grouping for segmentation; segment clustering is most useful for type clustering; it may also be used for composite objects, if used at a sufficiently course grain.

The largest performance gains are generally afforded by page clustering, since pages are the unit of access from the disk and a "working set" of pages is selected dynamically according to the access characteristics of an application program. Segment clustering produces efficiency gains only if relatively large contiguous units are transferred from disk, or when efficiency gains can be made through grouping operations (for example, for composite objects deletion).

**2.4. Users' hints**

To expedite the retrieval of related data, database systems often take hints from the user (or database administrator) to store related data physically close together [KIM90b]. For example, the GemStone database administrator, or a savvy application programmer can hint GemStone that certain objects are often used together and so should be clustered on disk [MAIE86]. The VBASE system allows explicit clustering hints when objects are created [ANDR91a]. The strategy adopted in ONTOS is to allow the programmer to specify clustering and to provide tools for reclustering when more experience with the applications permits better choices to be made [ANDR91b].

**2.5. Static versus dynamic clustering**

In the *static* case, clustering is done at the time objects are created and no reorganization is implied when the links between objects are updated [BENZ90]. A static clustering scheme offers a good placement policy for complex objects but does not take into account the dynamic evolution of objects. In applications such as design databases, objects are constantly updated during early parts of the design cycle. Frequent updates may destroy the initially clustered structure. To keep the object structure optimized, reorganization might be necessary for efficient future accesses [DEUX90].

*Dynamic* clustering is done at run time when objects are accessed concurrently and becomes attractive in an environment where the read operations dominate the write operations [BENZ90]. A dynamic clustering scheme should try to recluster when scattered access cost becomes too high. However, reclustering will generate overhead such as extra disk I/Os, so it is important to determine when a reorganization should occur. If the overhead is not justified, reclustering may actually degrade the overall performance [CHEN91].

# 3. CLUSTERING ALGORITHMS

## 3.1. Cactis clustering algorithm

*3.1.1. Algorithm presentation*

Cactis [HUDS89] is an object-oriented, multi-user DBMS developed at the University of Colorado. It is designed to support applications that require rich data modeling capabilities and the ability to specify functionally-defined data.

The Cactis clustering algorithm is designed to place objects that are frequently referenced together into the same block (i.e., page, i.e., I/O unit) on secondary storage. It can improve response time up to 60%.

In order to improve the locality of data references, data is clustered on the basis of usage patterns. A count of the total number of times each object in the database is accessed is kept, as well as the number of times each relationship between objects in the process of attribute evaluation or marking out-of-date is crossed. Then, the database is periodically reorganized on the basis of this information. The database is packed into blocks using the greedy algorithm shown in Figure 2.

---

**Repeat**
    Choose the most referenced object in the database that has not yet been assigned a block.
    Place this object into a new block.
    **Repeat**
        Choose the relationship belonging to some object assigned to the block such that:
            (1) the relationship is connected to an unassigned object outside the block and,
            (2) the total usage count for the relationship is the highest.
        Assign the object attached to this relationship to the block.
    **Until** the block is full.
**Until** all objects are assigned blocks.

---

Figure 2: Cactis clustering algorithm [HUDS89]

This clustering algorithm is also implemented in the Zeitgeist system [FORD88].

The Cactis clustering algorithm is a static algorithm since it is periodically used to recluster the database when the database is idle. This implies that the database is not clustered on the first run because no information about the database is available [CHAB93].

This algorithm does not require users' hints. This is an advantage since no arbitrary choice has to be made by the user [CHAB93]. But it also implies some time overhead (time to compute total number of times each object is accessed and number of times each relationship is crossed) and space overhead (the main memory space used to store the counters grows with the database size). It also raises the problem of getting pertinent statistics about the database.

*3.1.2. Clustering example with Cactis*

Let us say we want to cluster six objects into blocks of size 10. The objects' characteristics are given by Table 1. It gives for each object its size, the number of times it has been accessed, a list of objects with which it is related and the number of times each of these relationships has been crossed.

| Object name | Size | Number of times accessed | Relationships | Number of times crossed |
|---|---|---|---|---|
| O1 | 7 | 90 | O3 | 30 |
|    |   |    | O4 | 80 |
| O2 | 2 | 200 | O3 | 70 |
|    |   |    | O6 | 200 |
| O3 | 5 | 80 | O1 | 30 |
|    |   |    | O2 | 70 |
| O4 | 6 | 50 | O1 | 80 |
|    |   |    | O5 | 100 |
| O5 | 4 | 300 | O4 | 100 |
|    |   |    | O6 | 50 |
| O6 | 3 | 170 | O2 | 200 |
|    |   |    | O5 | 50 |

Table 1: Objects' characteristics for the clustering example with Cactis

Algorithm trace:  NEW BLOCK    O5 selected
                               O5-O4 relationship selected, O4 selected, block full
                  NEW BLOCK    O2 selected
                               O2-O6 relationship selected, O6 selected
                               O2-O3 relationship selected, O3 selected, block full
                  NEW BLOCK    O1 selected, all objects clustered

**3.2. ORION clustering method**

ORION is a series of next-generation database systems that have been prototyped at MCC (Microelectronics Computer Technology Corp.) as vehicles for research into the next-generation database architecture and into the integration of programming languages and databases [KIM90a]. ORION has been designed for Artificial Intelligence (AI), Computer-Aided Design and Manufacturing (CAD/CAM) and Office Information System (IOS) applications [BANE87].

ORION supports only a simple clustering scheme. Instances of the same class are clustered in the same physical segment (i.e., a number of blocks or pages). Each class is associated with one single segment. [KIM90a]

But ORION also provides direct support for *composite objects*, i.e., objects with a hierarchy of exclusive component objects (Figure 3). The hierarchy of classes to which the objects belong is a *composite object hierarchy*. The object-oriented data model, in its conventional form, is sufficient to represent a collection of related objects. However, it does not capture the IS-PART-OF relationship between objects; one object simply references, but does not own, other objects. A composite object hierarchy captures the IS-PART-OF relationship between a parent class and its component classes, whereas a class hierarchy represents the IS-A relationship between a superclass and its subclasses [BANE87].

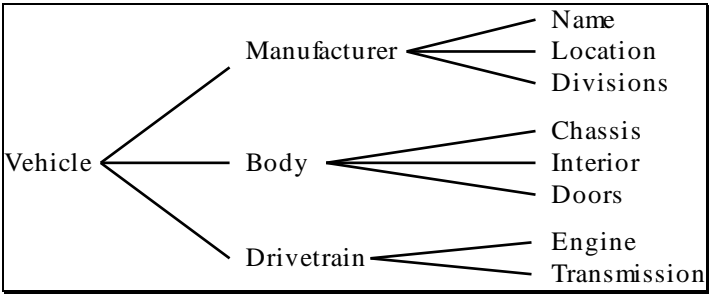

Figure 3: Example of composite object

Then it becomes advantageous to store instances of multiple classes in the same segment. User assistance is required to determine which classes should share the segment. The user can at run time issue a Cluster message containing a "ListOfClassNames" argument specifying the classes that are to be placed in the same segment [BANE87].

In ORION, segments have a fixed size. So the number of pages they contain gives the number of I/Os necessary to load the segment. When a segment is full, a new page is allocated and linked to the segment (a pointer must be maintained in the segment descriptor). This implies some overhead to find the address of each additional page [CHAB93].

The advantage of this method is its simplicity that makes the method fast and easy to implement since no cost model is defined and no overhead is implied to determine what is the optimal storage unit for an object. But simplicity also turns to a limitation since users' hints can only be based on the static information given by the data model and not on some information determined by the database usage and which could lead to a better clustering [CHAB93].

Figure 4 illustrates the way we implemented the ORION clustering method in our simulation models.

---

Considering a set of objects to cluster:

**1)** Select the first object to cluster.
**2)** Get the whole composite hierarchy (if any) attached to this object.
**3)** Cluster all the objects belonging to this composite hierarchy into a new segment.
**4)** Remove the objects belonging to this composite hierarchy from the set of objects to cluster.
**5)** Select the next object to cluster.
**6)** Reiterate from Step 2 until all the composite hierarchies are clustered.
**7)** Cluster together by class the remaining objects into distinct segments.

Figure 4: Our implementation of the ORION clustering method

### 3.3. CK clustering algorithm

The CK algorithm (from its authors' names: Chang and Katz) is defined in the CAD/CAM context. It can improve response time up to 200% when the Read/Write ratio is high (which is true for real CAD applications) [CHAN89b]. The CK algorithm makes use of several new concepts, such as structural relationships and instance-to-instance inheritance.

*3.3.1. Structural relationships*

Structural relationships are versions, configurations and equivalence relationships.

Objects sharing the same interface but having different implementations are called versions [BATO85]. They represent different design alternatives. For example, if an object is identified by the pattern: *Name[Version].Type* where "Name" is the object name, "Version" its version number and "Type" its type; Nice[1].car, Nice[2].car and Nice[3].car would be three versions of the same object "Nice" type of which is "car".

A very important characteristic of OODBs is the presence of composite (complex or nested) objects. This concept is represented through composite/component relationships among objects. Coupling the concept of versions with composite objects leads to configurations. A configuration is a composite unit whose components are bound to specific versions (Figure 5) [CHAN90].

If two objects are alternative representations of the same real world entity, they are equivalent.

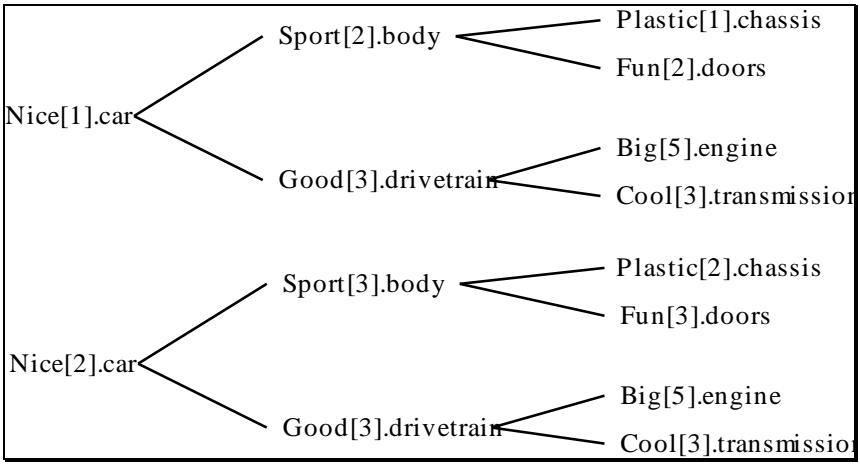

Figure 5: Example of configurations

*3.3.2. Instance-to-instance inheritance*

Besides structural relationships, inheritance provides additional semantics. As in object-oriented programming languages, a class/subclass hierarchy can be defined for an OODB based on the IS-A relationship. A subclass inherits the structure (i.e., attributes' definitions) and the methods of its superclass. However, in OODBs, this form of inheritance (called type inheritance) is not sufficient. [CHAB93]

The CK algorithm also uses instance-to-instance inheritance that not only transfers the existence of attributes from one object to another (like type inheritance), but moreover the values of these attributes [WILK88].

Instance-to-instance inheritance is important in computer-aided design databases, since a new version tends to resemble its immediate ancestor. It is useful if a new version can inherit its attributes' values, and more importantly its constraints, from its ancestor [KATZ91].

*3.3.3. Algorithm presentation*

Instance-to-instance inheritance introduces more complexity because it allows attributes to be selectively inherited at run-time. This run-time flexibility requires a sophisticated approach for clustering. The CK algorithm is based on inter-objects access frequencies (given by the user at data type creation time) for each kind of structural relationship, e.g., 20% of access along version relationships, 75% of access along configuration relationships and 5% of access along equivalence relationships.

When a new object is created, the algorithm chooses an initial placement based on which relationship is most frequently used to reach the object (in the above example, a new instance would probably be placed in the same page as its composite objects). Then, for each inherited attribute, cost formulas are used to choose between implementation by copy or by reference, i.e., either by copying the attribute's value or referencing it with a pointer. The augmented access frequencies (i.e., relationship traversal frequencies plus inheritance traversal frequencies) may change the initial placement. The clustering algorithm pseudo code is given in Figure 6.

Then, if the best candidate page is full, either the next best candidate page is chosen or the page is split if the expected access cost resulting from the split is an improvement over placement in the next best candidate page.

Page splitting is performed by a greedy algorithm that partitions the inheritance-dependency graph into two sub-graphs that each fit into one page. This algorithm is not optimal, but it is linear (whereas an exact partitioning algorithm would be NP-complete). It is described in Figure 7.

- <u>Step 1</u>: get initial information (Is page splitting enabled? What are the attributes implemented by reference?, etc.)
- <u>Step 2</u>: calculate lookup cost for attributes implemented by reference for each page
    FOR each page
        FOR each attribute implemented by reference
            IF attribute NOT IN page THEN increase lookup cost for the page
- <u>Step 3</u>: calculate lookup and storage costs for attributes implemented by copy for each page
    FOR each attribute implemented by copy
        FOR each page
            IF attribute NOT IN page THEN increase lookup and storage costs for the page
- <u>Step 4</u>: calculate total cost of every page
- <u>Step 5</u>: pick up the best candidate page and try to insert the object
    candidate page = page with lower total cost
    IF cluster policy IS no split THEN
        WHILE object does not fit into candidate page DO candidate page = page with next lower total cost
    IF cluster policy IS page split AND object does not fit into candidate page THEN
        page_split (candidate page)

Figure 6: CK clustering algorithm

The Page_split algorithm assumes that the arc costs $C_{e_i}$ (i.e., run-time lookup cost) between objects are always maintained and sorted. The node capacity $Cap_{v_i}$ (i.e., the object size) is also maintained. Subset A and B represent the sets of objects assigned to the new pages after splitting. Both subsets are empty ate the beginning. E is the initial set of arcs relating the objects.

- Step (1): Select the maximum value arc from E as $e_{target}$ and set E to be (E - {$e_{target}$}). Let $v_{head}$ and $v_{tail}$ be the head and the tail nodes of $e_{target}$.
- Step (2): Supposed both $v_{head}$ and $v_{tail}$ are new to subsets A and B. Insert $v_{head}$ and $v_{tail}$ in subset A if $Cap_{v_{head}}$ plus $Cap_{v_{tail}}$ is less than the remaining capacity of subset A. Otherwise, insert $v_{head}$ and $v_{tail}$ in subset B if subset B has space for these nodes. If neither subset A or B could accommodate both $v_{head}$ and $v_{tail}$, a broken arc is found and $C_{e_{target}}$ is added into $C_{total}$.
- Step (3): Supposed $v_{head}$ is in subset A and $v_{tail}$ is not in subset A or B. Insert $v_{tail}$ into subset A if feasible. Otherwise, a broken arc is found and $C_{e_{target}}$ is added into $C_{total}$.
- Step (4): Supposed both $v_{head}$ and $v_{tail}$ are visited before, a broken arc is found and $C_{e_{target}}$ is added into $C_{total}$.
- Step (5): Look back to step (1) until arc set E is empty.

Figure 7: Page_split algorithm

### 3.3.4. Clustering example with CK

Let us consider the object hierarchy given by Figure 8. Objects are represented according to the following format: *Name[Version].Type* where "Name" is the object name, "Version" its version number and "Type" its type. Numbers above arcs represent the run-time look-up cost of the structural relationship. We want to cluster these objects in pages of size 5. Table 2 gives types' characteristics. Objects are clustered in their creation order. The algorithm trace is provided by Table 3.

| Type | Object size | Access frequency along version | Access frequency along configuration | Access frequency along equivalence |
|---|---|---|---|---|
| Ferrari | 2 | 20% | 10% | 70% |
| car | 2 | 65% | 30% | 5% |
| body | 3 | 25% | 75% | 0% |
| drivetrain | 3 | 30% | 70% | 0% |

Table 2: Types characteristics for the clustering example with CK

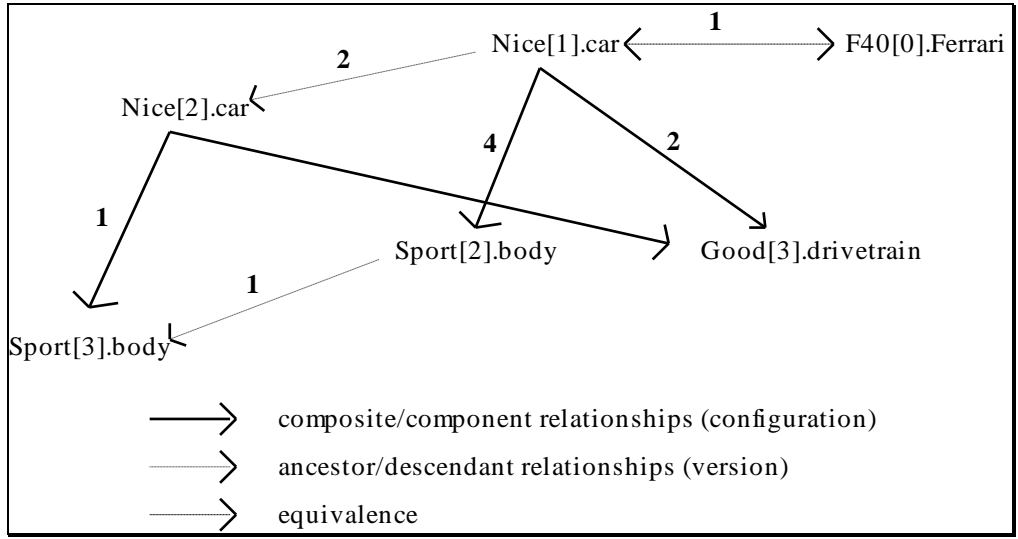

Figure 8: Sample object hierarchy

| Object to cluster | Placement | | | Notes |
|---|---|---|---|---|
| | along relationship | with object | in page | |
| Nice[1].car | version | — | #1 | — |
| F40[0].Ferrari | equivalence | Nice[1].car | #1 | — |
| Sport[2].body | configuration | Nice[1].car | #1: impossible | – *No Page Splitting:* placement in page #2<br>– *Page Splitting:* Nice[1].car and Sport[2].body are placed together in page #1, F40[0].Ferrari is moved to page #2 |
| Good[3].drivetrain | configuration | Nice[1].car | #1: impossible | – *No Page Splitting:* placement in page #3<br>– *Page Splitting:* Nice[1].car and Sport[2].body stay together in page #1, Good[3].drivetrain is placed into page #3 |
| Nice[2].car | version | Nice[1].car | #1: impossible | – *No Page Splitting:* placement in page #4<br>– *Page Splitting:* Nice[1].car and Sport[2].body stay together in page #1, Nice[2].car is placed into page #4 |
| Sport[3].body | configuration | Nice[2].car | #4 | — |

Table 3: CK algorithm trace

## 4. SIMULATION MODEL

We choose to use simulation to compare the Cactis clustering algorithm, the CK clustering algorithm and the ORION clustering method for several reasons. First of all, it is important for the results to be meaningful that the performance evaluation is done using the same "environment" for each algorithm, since we focus specifically on clustering. Therefore we could not have benchmarked each OODB since Cactis and ORION use, for example, different buffering and caching strategies. Furthermore, CK clustering algorithm is not implemented in an OODB yet. Building our own simulation model allows us to ensure that the algorithms are compared in the same conditions. Mathematical analysis has also already been performed on these three algorithms [CHAB93]. Although it provides exact results, it only gives a general idea of the algorithm performances and cannot detect in which specific cases an algorithm performs better than an other as simulation can.

### 4.1. Object base

For our simulations, we used a random object base whose class hierarchy forms a DAG, as in [HE93]. The database generation was performed in two phases: first generate class hierarchies and class definition (Figure 9), then generate instances for these classes. To simplify the class hierarchy, we did not take into account multiple inheritance because it has no effect on clustering. We also assumed that a given class had one single ancestor version and one single descendant version but could have several component classes or equivalent classes.

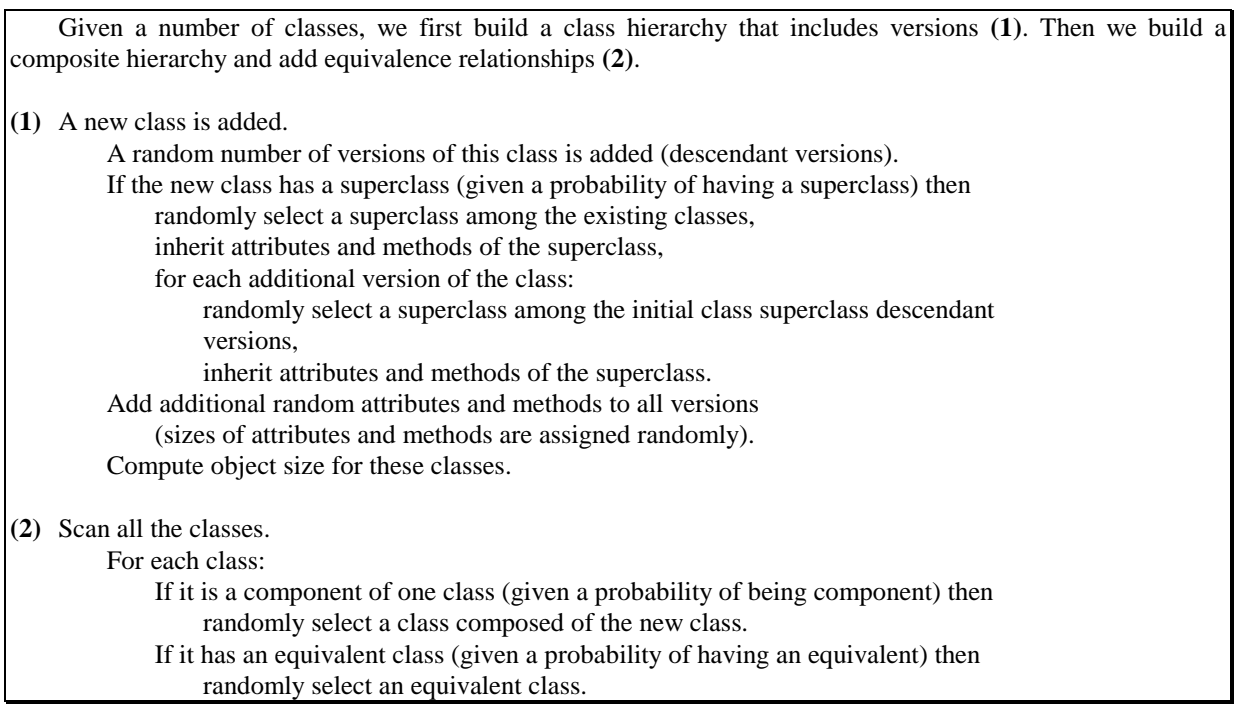

Given a number of classes, we first build a class hierarchy that includes versions (**1**). Then we build a composite hierarchy and add equivalence relationships (**2**).

(**1**) A new class is added.
    A random number of versions of this class is added (descendant versions).
    If the new class has a superclass (given a probability of having a superclass) then
        randomly select a superclass among the existing classes,
        inherit attributes and methods of the superclass,
        for each additional version of the class:
            randomly select a superclass among the initial class superclass descendant
            versions,
            inherit attributes and methods of the superclass.
    Add additional random attributes and methods to all versions
        (sizes of attributes and methods are assigned randomly).
    Compute object size for these classes.

(**2**) Scan all the classes.
    For each class:
        If it is a component of one class (given a probability of being component) then
            randomly select a class composed of the new class.
        If it has an equivalent class (given a probability of having an equivalent) then
            randomly select an equivalent class.

Figure 9: Class lattice generation

Instance creation has been designed as a special kind of transaction. However, the initial database is to be created before any other query can occur, given an initial number of objects. The method we used to generate instances is shown in Figure 10.

```
For each new object:
    Randomly select a class.
    If the new object class is a component of another class then
        randomly select an instance of this class (if any) to be composed of the new object.
    If the new object class is a version then
        randomly select one ancestor object in the new object class ancestor class,
        If using CK, inherit values of common attributes (either by copy or by reference).
    If the new object class has an equivalent class then
        randomly select one equivalent object among instances of the equivalent class.
```

Figure 10: Instances generation

### 4.2. Query generation

The HyperModel Benchmark [ANDE90, BERR91] provides 20 different types of transactions. From those 20, we have isolated 15 types of transactions (some of them are slightly modified to match the structural relationships we use) that were relevant to study the behavior of clustering algorithms. Each transaction has a probability to occur.

- *Name Lookup*: Retrieve a randomly selected object; fetch one of its (randomly selected) attributes' value.
- *Range Lookup*: Select a class at random; select one of its attributes at random; determine randomly two test values; fetch all the attributes of all the instances of the class whose selected attribute's value are in the range defined by the test values.
- *Group Lookup*: Given a randomly selected starting object, fetch all the attributes of either:
  - all its component objects,
  - all its equivalent objects,
  - all its descendant versions.
- *Reference Lookup*: Given a randomly selected starting object, fetch all the attributes of either:
  - its composite object,
  - all its ancestor versions.
- *Sequential Scan*: Select a class at random; select one of its attributes at random; fetch this attribute's values for every instance of the class.
- *Closure Traversal*: Given a randomly selected starting object, follow one of the three structural relationships (i.e., version, configuration or equivalence) to a certain predefined (random) depth D; fetch a random attribute from the resulting object; the followed relationship can be either always the same or randomly selected.
- *Editing*: Select an object at random; update one of its attribute (randomly chosen) with a random value.

- *Object Creation*: Creation of a new object (cf. object base generation). This activates the CK clustering algorithm.
- *Reclustering*: The ORION clustering algorithm needs a "Cluster message" to be activated at run time [BANE87]. The Cactis clustering algorithm is static. We can assume it will also wait for a cluster message before reorganizing the database. However, cluster messages for the Cactis algorithm should be far less frequent than cluster messages for the ORION algorithm since the Cactis clustering algorithm is supposed to run when the database is idle [HUDS89].

**4.3. Overall model**

The overall simulation model is inspired by the one provided in [CHAN89a]. It is composed as follows (Figure 11).

- <u>Client module</u>: After a predefined think time, the client issues the transactions to the Transaction Manager according to some frequencies of occurrence.
- <u>Transaction Manager module</u>: The transaction manager extracts from transactions which objects have to be accessed or updated, and performs the operations. In the case of a regular operation, object requests are sent to the Buffering Manager. In the case of instance creation or a Cluster message, the Clustering Manager is invoked.
- <u>Buffering Manager</u>: The Buffering Manager checks if an object is in main memory and requests it to the I/O Subsystem if it is not. It also deals with the page replacement strategy (when a new page is needed, the oldest page in memory is dropped and replaced by the new one; if the dropped page has been modified, it is saved on secondary storage).
- <u>Clustering Manager</u>: The Clustering Manager is activated depending on the algorithm (i.e., Cactis, CK or ORION) it implements. It deals with reorganizing the database on secondary storage to achieve better performance.
- <u>I/O Subsystem</u>: This module deals with physical accesses to secondary storage.

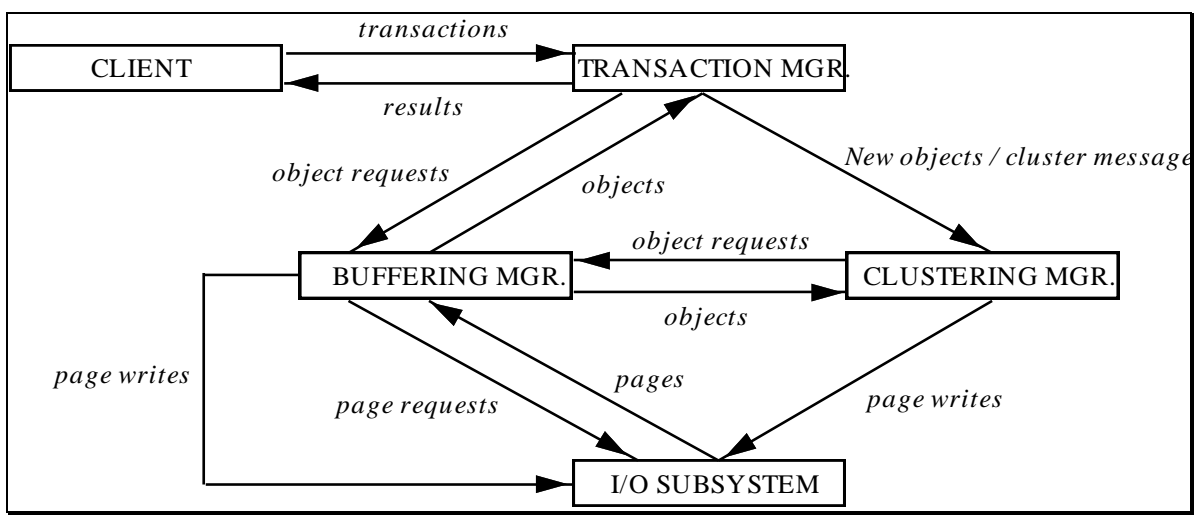

Figure 11: Overall simulation model

### 4.4. Simulation parameters

Parameters are divided into two categories: static parameters that may not change from one simulation to another and dynamic parameters that can vary from one simulation to another. Tables 4 and 5 provide the simulation parameters we used for our simulation experiments.

| Parameter name | Designation | Value | References |
|---|---|---|---|
| RCC | Average locking/unlocking time (concurrency control) | 0.5 ms | [SRIN91] |
| IMLVL | Multiprogramming level | 10 | [GRUE91] |
| IWDSIZE | Memory word size | 4 bytes | [GRUE91] |
| ICPU | CPU power | 2 Mips | [GRUE91] |
| RMACC | Memory word access time | 0.0001 ms | [GRUE91] |
| RMTEST | Time for comparison of two memory words | 0.0007 ms | Two memory accesses, one subtraction |
| IPGSIZE | Size of disk page | 2048 bytes | [CHEN91] |
| RSEEK | Average disk seek time | 28 ms | [CHEN91] |
| RLATENCY | Average disk latency time | 8,33 ms | [CHEN91] |
| RTRANSFER | Disk page transfer time | 1.28 ms | [CHEN91] |

Table 4: Static parameters

Note: We chose to use a relatively small page size in our simulations. Hence, to remain consistent in terms of scale, we also chose to use small objects. These choices have been made for two reasons:
- for simplicity,
- to minimize the size of the database in main memory during the simulations. Since simulation is very greedy in terms of memory, it is important to optimize memory usage.

### 5. SIMULATION RESULTS

To compare the performance of the three clustering algorithms, we conducted three testing cases: varying the database size, the buffer capacity and the Read/Write ratio.

The comparison criteria we adopted are the following:
- *Response Time:* response time is measured for all transaction types except reclustering (which is considered as a special transaction in the Cactis and ORION simulation models; however, time when transactions are blocked because of a reclustering is also taken into account); it is a good metric for overall performance;
- *Transactions I/Os:* transactions I/Os is the number of I/Os performed to complete regular transactions; transactions I/Os may be an indication on how well objects are clustered;

| Parameter name | Designation | Default value | Range |
| --- | --- | --- | --- |
| RAVGTHINK | Average client think time | 4 s | 0.1-10 s |
| NCL | Number of classes | 20 | 10-30 |
| IAVGVER | Average number of versions per class | 3 | 1-5 |
| RPSUPER | Probability for a class of having a superclass | 0.9 | 0-1 |
| RPCOMP | Probability for a class of being a component class | 0.5 | 0-1 |
| RPEQUI | Probability for a class of having an equivalent class | 0.1 | 0-1 |
| INOBJ | Initial number of objects | 400 | 100-1000 |
| IAVGASIZE | Average attribute size | 1 word | 1-3 words |
| IAVGNATTR | Average number of attributes per class | 10 | 5-20 |
| IBUFF | Size of memory buffer | 10 pages | 10-100 pages |
| IMD | Maximum depth in Closure Traversals | 5 | 3-10 |
| ISEGSIZE | Default segment size (ORION) | 5 | 3-10 |
| ITHRESHOLD | Update Threshold (CK) | 25 | 0-255 |
| ISCALEF | Scale factor (CK) | 0.5 | 0-1 |
| ISPLIT | Page split policy (CK) | ON | ON/OFF |
| PT1-PT12 | Probability of Read Transaction (#1-12) | 0.065 | 0-1 |
| PT13 | Probability of Editing | 0.1695 (Cactis) 0.169 (ORION) 0.17 (CK) | 0-1 |
| PT14 | Probability of Object Creation | 0.05 | 0-1 |
| PT15 | Probability of Reclustering | 0.0005 (Cactis) 0.001 (ORION) 0 (CK) | 0-1 |
| SIMTIME | Simulation Time | 10,800 s | 3,600-86,400 |

Table 5: Dynamic parameters

- *Clustering Time Overhead:* clustering time overhead measures the time needed to reorganize the database; it includes I/O time and the time necessary to perform the memory operations needed by the clustering algorithm but it does not take into account the counters updates performed by Cactis and CK since those take a negligible amount of time compared to even one single I/O;
- *Clustering I/O Overhead:* clustering I/O overhead is the number of I/Os performed during database reorganizations and object clustering;
- *Maximum number of pages used:* maximum number of pages used is the maximum number of disk pages needed by a clustering algorithm to cluster all the objects of the database;
- *System Throughput:* system throughput is the number of transactions completed per second.

Due to space limitation, we present in the following subsections only some of the results we obtained.

## 5.1. Effects of the database size

We first tested the effect of varying the initial number of objects in the database using a uniform random distribution to choose the transactions' starting objects. This is not always realistic since there may be objects that are "hotter" (i.e., more frequently accessed) than others in real world applications. Furthermore, the performance of the Cactis clustering algorithm depends on run-time computed statistics, such as object's access frequencies, that are not the same when using different random distributions to select the transactions' starting objects. So we implemented in a second series of simulations a normal random distribution for the transactions starting object, which is very similar to the "skewed random" function introduced in [TSAN92a].

Figure 12a shows that response time when using Cactis is 24% lower than when using ORION. Figure 12b shows that, for CK, response time increases linearly with the number of objects and that CK algorithm totally outperforms the other two (being about 800 times better than Cactis). In the CK graphs, "PS ON" and "PS OFF" stand for page splitting enabled and page splitting disabled. "U" stands for uniform random distribution for starting object and "N" stands for normal random distribution for starting object.

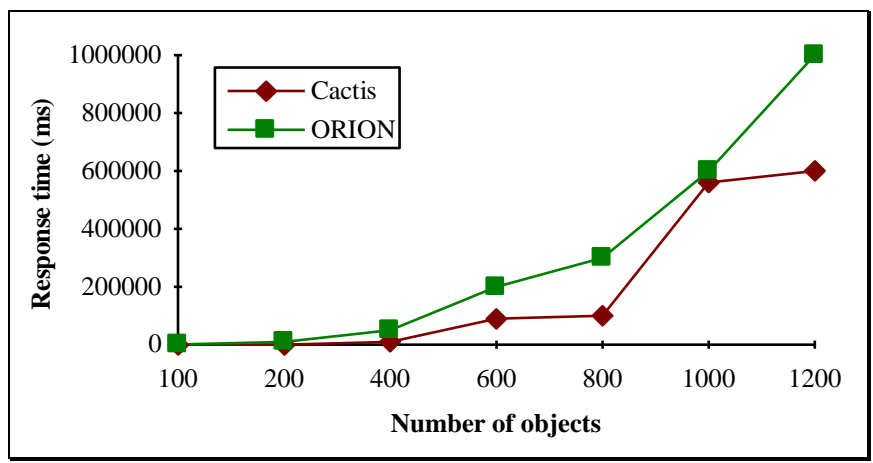

Figure 12a: Response time function of number of objects (U)

Figures 12a and 13 show indeed that Cactis performs 1.5 times better when objects are not accessed through a uniform distribution, especially for intermediate numbers of objects (between 400 and 600).

On the contrary, the CK and ORION algorithms does not show any significant change of performance since they do not use such statistics.

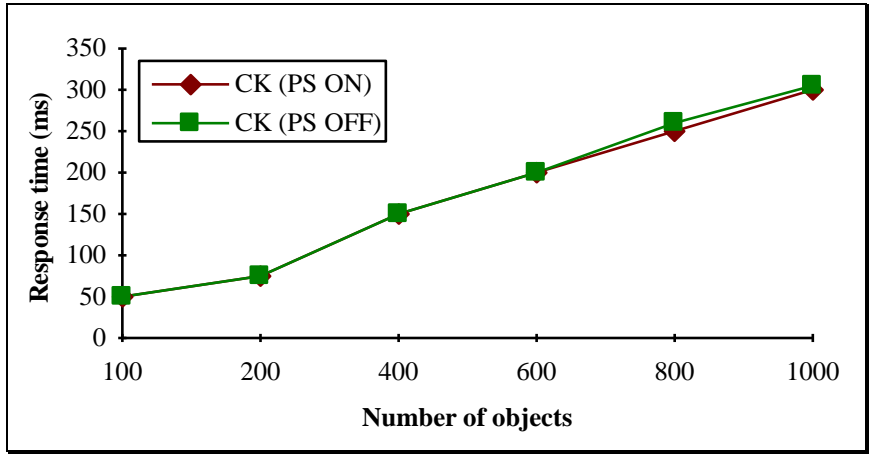

Figure 12b: Response time function of number of objects (U)

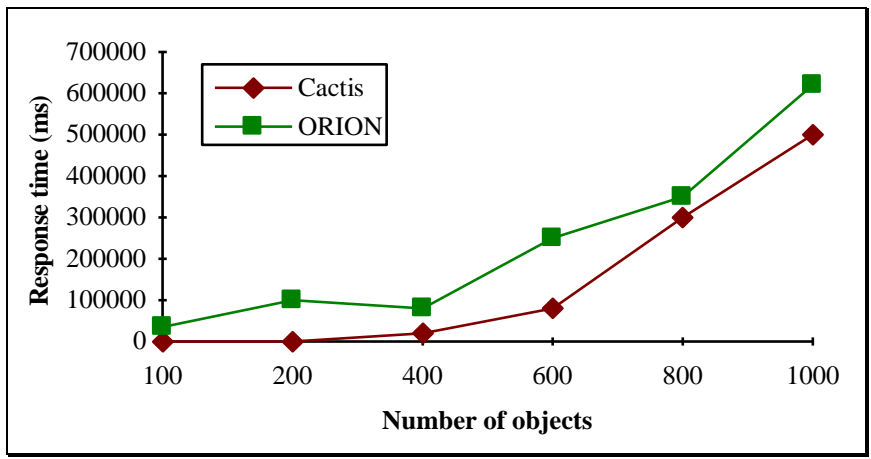

Figure 13: Response time function of number of objects (N)

Transactions I/Os give an idea of how good is a clustering scheme. Figure 14 thus shows that objects are 3.8 times better clustered by Cactis and CK (Cactis being slightly better) than they are by ORION. It seems surprising that Cactis clusters so well and performs worse than CK, but Figures 15a and 15b show again that CK outperforms both Cactis and ORION in terms of low clustering overhead (being 266 times better than Cactis and 502 times better than ORION). Furthermore, clustering overhead is almost constant for CK. Such an outstanding performance is due to the true dynamic nature of CK, which is called only at object creation time and only accesses the object to cluster related objects once. Variations in clustering overhead come from variations in the number of created objects.

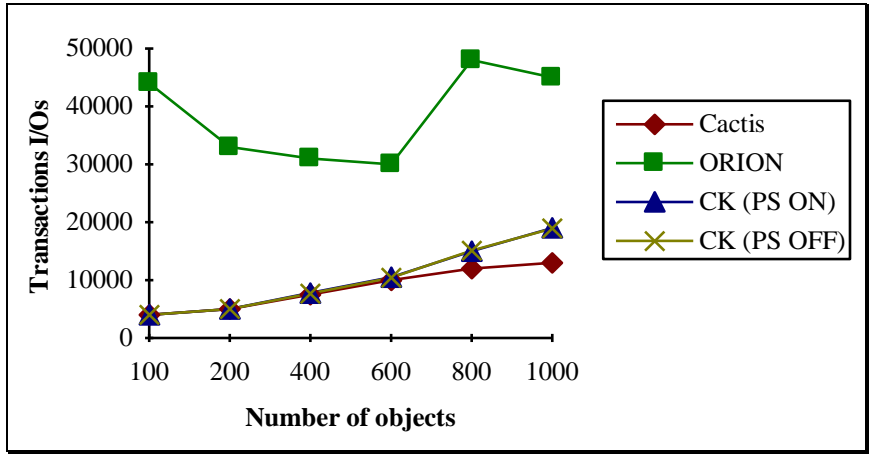

Figure 14: Transaction I/Os function of number of objects (U)

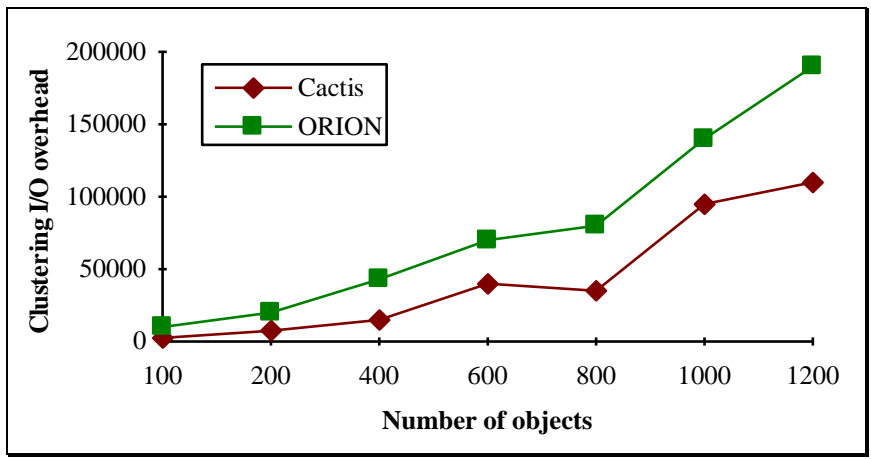

Figure 15a: Clustering I/O overhead function of number of objects (U)

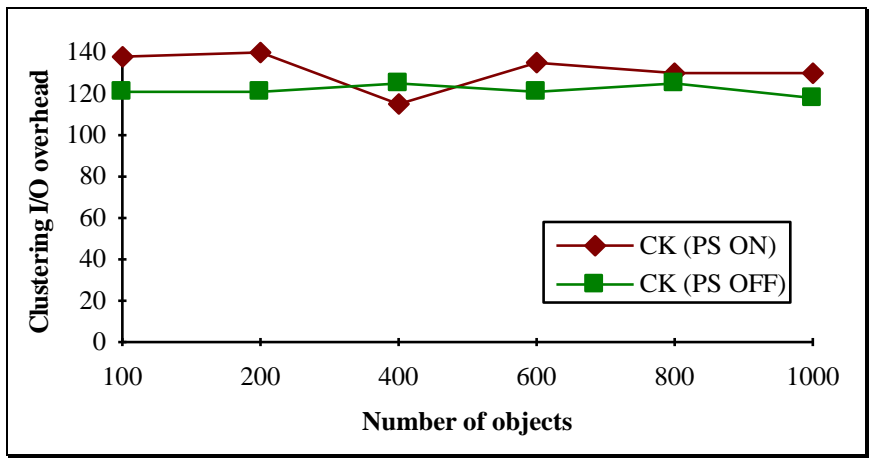

Figure 15b: Clustering I/O overhead function of number of objects (U)

The more a clustering algorithm is complex (i.e., the more it clusters object according to precise rules), the greater amount of disk pages it uses to cluster the object base. The maximum number of disk pages used (Figure 16), as expected, is higher for the more complex algorithms, i.e., CK needs 1.8 times as many pages as Cactis and Cactis needs 1.3 times as many pages as ORION, for which number of pages increases linearly.

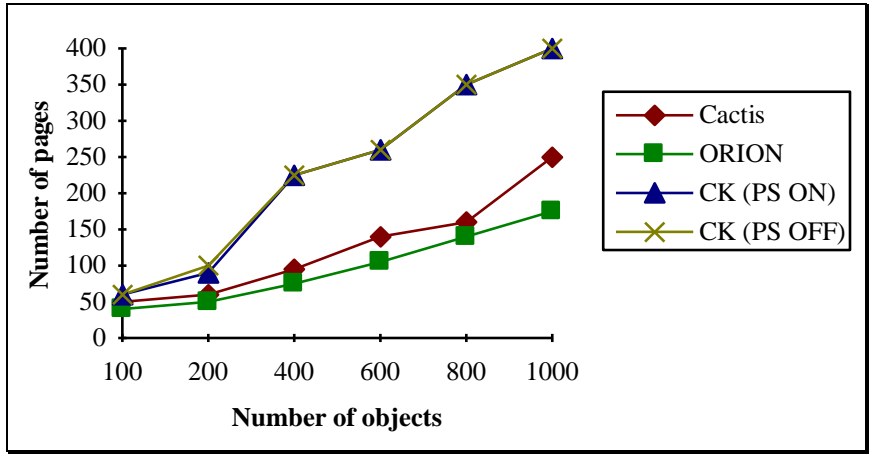

Figure 16: Maximum number of pages used function of number of objects (U)

Since average client think time (i.e., time between two transaction generations) is 4 seconds, optimal throughput lies around 0.25 transactions per second. Figure 17 is coherent with Figures 12a and 12b, showing a near constant throughput for CK. Throughput is high for all algorithms because a typical transaction is executed in much less time than the average think time.

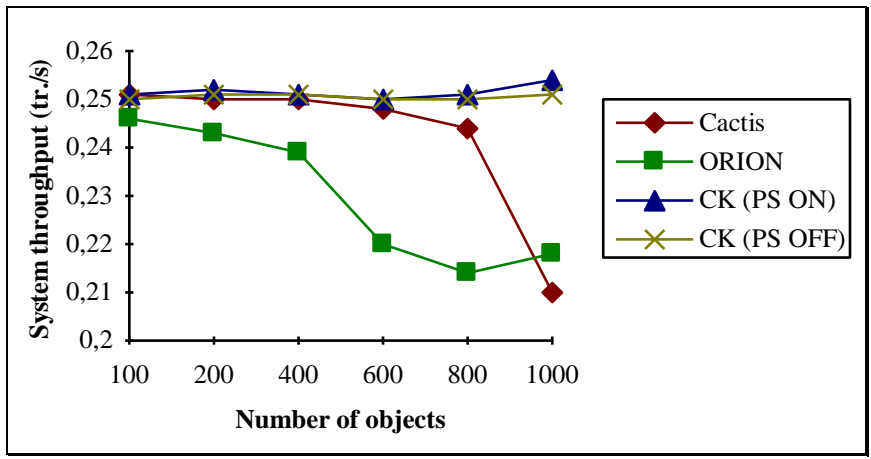

Figure 17: System throughput function of number of objects (U)

Note: Though page splitting should allow a better performance, our simulations show that the CK algorithm performances are very close whenever using the page splitting policy or not. This is due to our implementation of the page splitting algorithm that is not as efficient as it could be in reality, because we had no way to know or compute lookup costs in our simulations. Hence we used random lookup costs and thus could not achieve optimal object placement.

## 5.2. Effect of the buffer capacity

On one hand, increasing the buffer capacity may lessen the effects of clustering since a given set of related objects has a higher probability of being in main memory instead of on secondary storage. On the other hand, objects are also accessed when reorganizing the database. Thus, increasing buffer capacity should decrease clustering overhead, especially for the ORION clustering algorithm that may make several non-consecutive accesses to each object. We performed this set of simulations using an initial database of 400 objects and selecting the transactions' starting objects from a uniform random distribution.

Figures 18a and 18b show that response time decrease linearly with the buffer size for the Cactis and CK algorithms. The dual effect of increasing the buffer capacity is seen on both transactions I/Os and clustering overhead (Figures 19, 20a and 20b). The ORION algorithm has a similar behavior, but the gain in performance is seen earlier than with the other algorithms and then the gain in performance is less important (Figure 18a). We can explain this by the fact that the ORION algorithm uses a smaller amount of pages than the other algorithms to cluster the database. Thus, the buffer size grows faster relatively to the database size. For instance, a buffer size of 20 pages represents 25% of the database size for ORION versus only 15% for Cactis and 9% for CK. Figure 18a presents rather big variations in performance for ORION when buffer capacity grows over 40 pages. These odd variations are due to random events in the simulations. However, they oscillate around a mean value that decreases slowly but linearly.

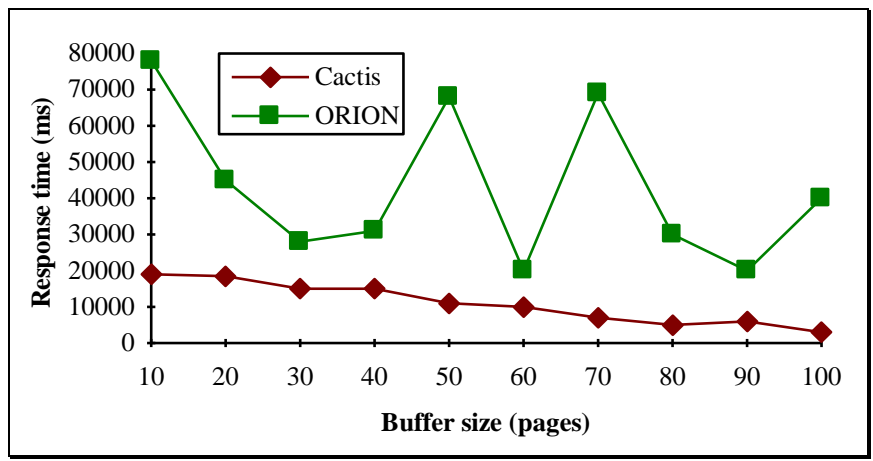

Figure 18a: Response time function of buffer size

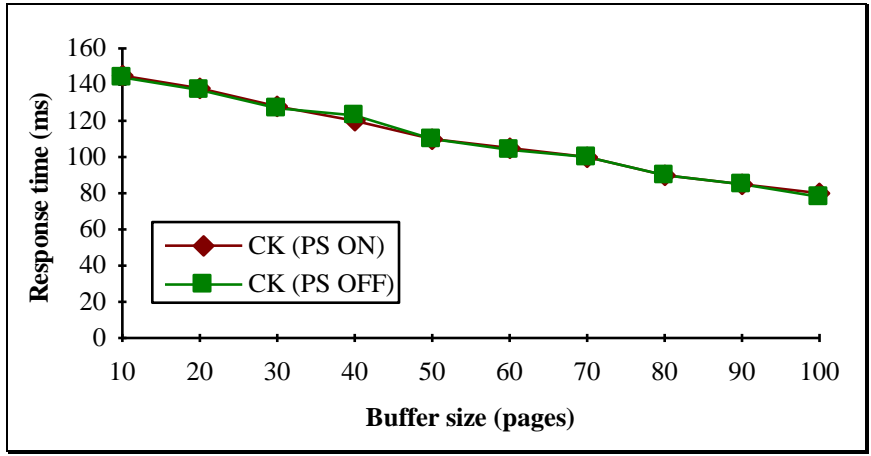

Figure 18b: Response time function of buffer size

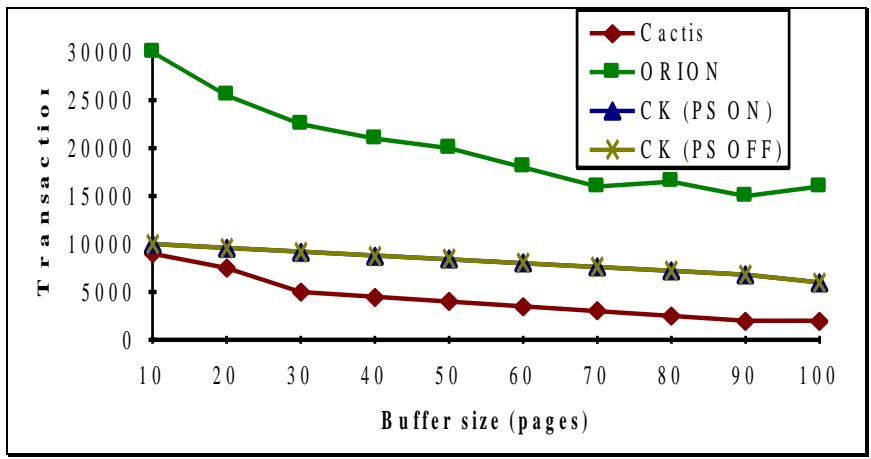

Figure 19: Transaction I/Os function of buffer size

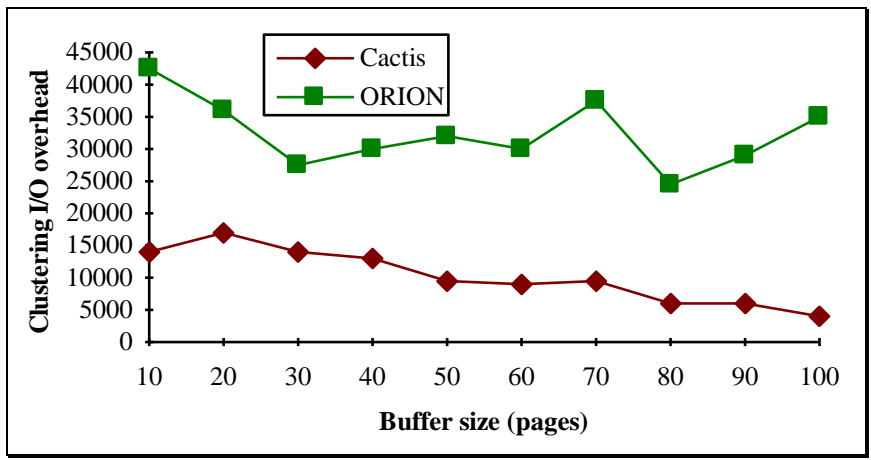

Figure 20a: Clustering I/O overhead function of number of buffer size

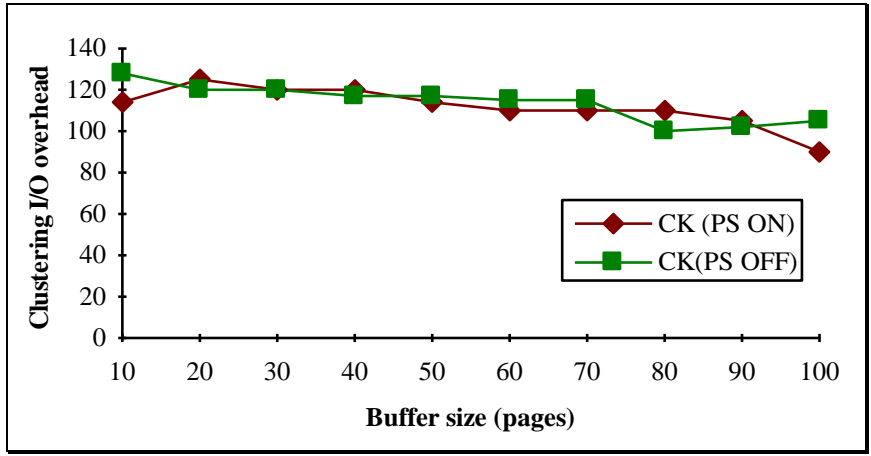

Figure 20b: Clustering I/O overhead function of number of buffer size

## 5.3. Effect of the Read/Write ratio

Read/Write ratio is an important factor when seeking to evaluate DBMSs performances. Furthermore, [CHAN89a] claims that CK algorithm performs better when the Read/Write ratio is high. For our simulation experiments, we used an initial database of 400 objects and a buffer size of 10 pages.

The performance of the Cactis and ORION algorithms decreases when the Read/Write ratio decreases (Figure 21a). On the contrary, response time decreases along with the Read/Write ratio in the case of CK (Figure 21b).

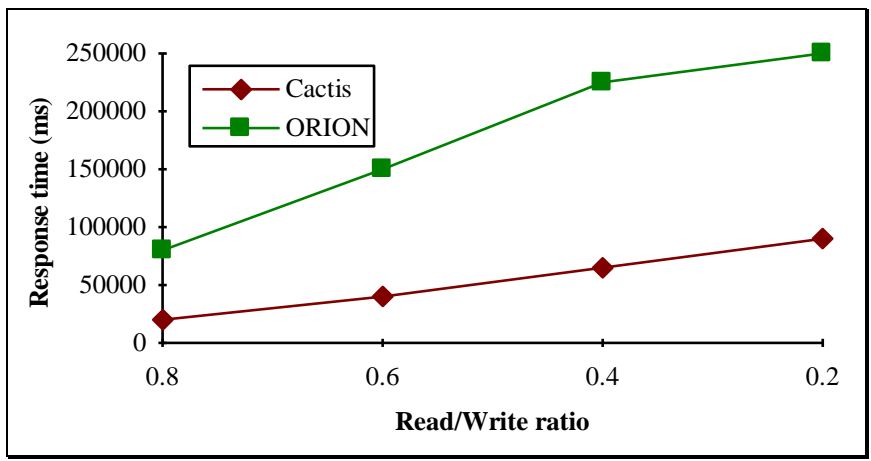

Figure 21a: Response time function of Read/Write ratio

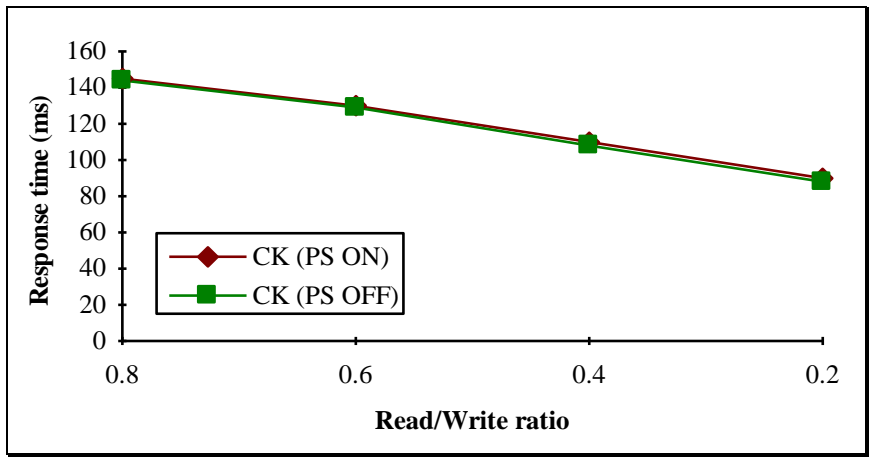

Figure 21b: Response time function of Read/Write ratio

Since Object Creation is a write operation, the more the Read Percentage drops, the more the database size increases, thus implying more clustering overhead and confirming what is said in [CHAN89a] (Figures 23a and 23b). Parallely, transactions I/Os are slowly decreasing in number for Cactis and CK (Figure 22). This is because one single Object Creation is less costly than, for instance, such read transactions as Sequential Scans or Range Lookups. That explains the raise in performance for CK, since transactions I/Os drops from 10,000 to 5000 while clustering I/O overhead only rises from 100 to 500. In the Cactis case, clustering overhead is too important to compensate the decrease in transactions I/Os. For ORION, transactions I/Os increase anyway because of the poor clustering ability of the algorithm (Figure 22).

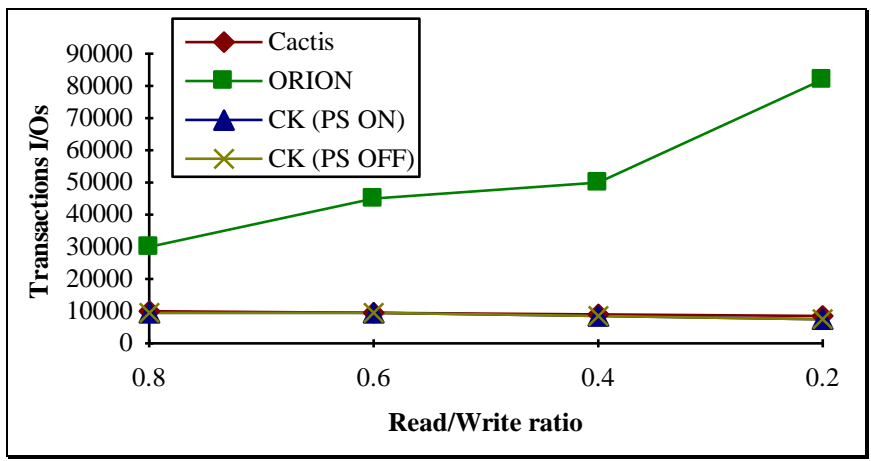

Figure 22: Transaction I/Os function of Read/Write ratio

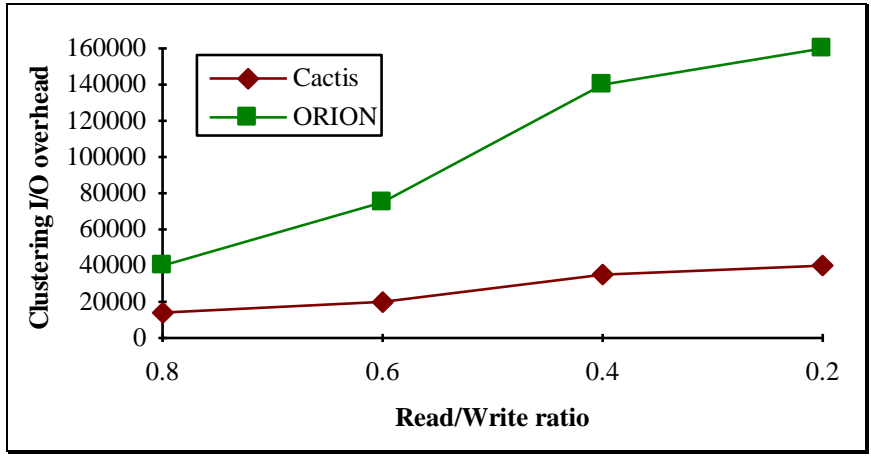

Figure 23a: Clustering I/O overhead function of Read/Write ratio

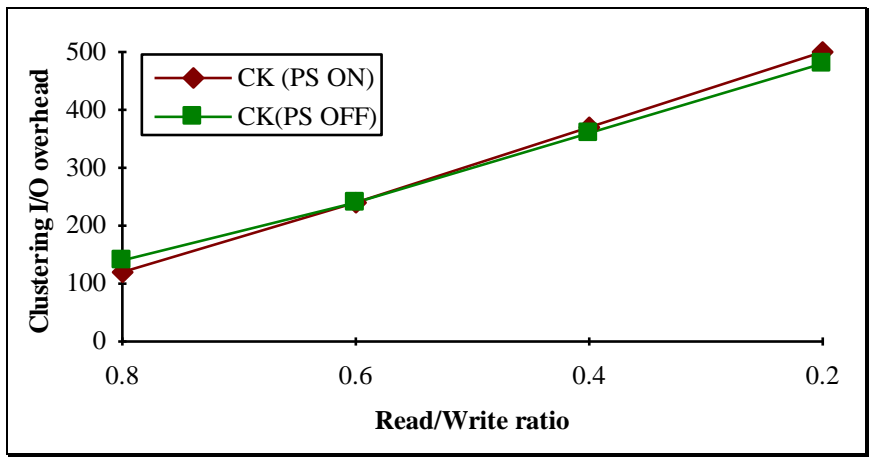

Figure 23b: Clustering I/O overhead function of Read/Write ratio

## 6. CONCLUSION

It is clear from our simulation experiments that the CK algorithm outperforms both Cactis and ORION in terms of overall performance. The results we obtained showed that this is due to both a good clustering capability and to the dynamic conception of the algorithm that allows an extremely low clustering overhead. Such a good behavior is achieved because the CK algorithm is activated only at object creation time and only accesses the few objects that are related to the newly created object once. Therefore, transactions are never blocked very long during clustering, as they are when the Cactis or the ORION algorithm are used. (The Cactis and ORION algorithms have to access all the objects in the database, even several times in the case of ORION, to reorganize the database; and transactions cannot be run when a reorganization occurs). CK's good clustering capability is based on the users' hints that specify the inter-objects access frequencies for each structural relationship and thus allows to cluster together objects that are likely to be accessed together.

Our simulations showed too that Cactis also had a good clustering capability. This is due to the use of statistics (i.e., objects access frequencies and relationships use frequencies) that allow to cluster together objects that are actually accessed together. However, the Cactis algorithm is still completely outperformed by the CK algorithm. This is because, when using Cactis, clustering overhead increases very quickly with the number of objects, thus annihilating any gain achieved from good clustering capability.

However, in order to maximize any gain from clustering, we made Cactis and ORION reorganize the database quite frequently. Hence, we obtained a high clustering overhead for these two algorithms. We have though to keep in mind that the Cactis algorithm has been designed to run when the database is idle, so that reclustering does not alter the database performance. Thus, if clustering overhead was not taken into account, the Cactis algorithm should perform about as well as the CK algorithm as long as the statistics used during the last reorganization are valid. Our simulations also show that ORION's clustering capability is much lower than that of both Cactis and CK. Hence, if the ORION clustering algorithm was run less frequently, even with a big gain from the decrease in clustering overhead, ORION would still not behave as well as CK. Furthermore, any gain from clustering achieved after these frequent reorganizations is then canceled, making the performances worsen.

In terms of disk space, the simpler a clustering algorithm is, the less space it should use. Actually, the more complex a clustering algorithm is, the more it clusters objects according to sharp criteria (for instance, not only according to the objects' classes, but also to various structural relationships, etc.). Thus, a smaller number of objects are likely to be clustered in the same clustering unit (either a page or a segment). So the number of pages needed to store the database is greater. Our simulation experiments confirm that fact. The ORION algorithm is the less greedy algorithm in terms of disk pages used. Then the Cactis algorithm follows, using almost half the number of disk pages needed by CK to cluster the database. However, when reorganizing the database, the Cactis and ORION algorithms need to build a new set of pages before deleting the old one. Thus they require about twice as much space as our graphs show.

Future research about this subject could be in a first step completing the simulation tests, notably by determining the effects of database changes, e.g., varying the probability for a class to belong to a composite class hierarchy, the average number and size of attributes, etc. We also need to perform tests when the workload varies (variation of the average client think time and the transactions' probabilities).

Then the next step would be the design and evaluation of one or several new clustering algorithms. One alternative would be to build a dynamic clustering algorithm that would use the same statistics as Cactis (i.e., objects access frequencies and relationships use frequencies) to cluster objects together, but could be able to use them at run-time (e.g., an algorithm activated at object creation time, like CK). Such an algorithm could have Cactis' good clustering capability without being handicapped by an important clustering overhead.

Another alternative would be to modify the CK algorithm so that it does not use users' hints for inter-objects access frequencies any more and rely only on statistics, in order to make the algorithm even more accurate and performant. The problem with users' hints is that their accuracy depends on the user's (either the database administrator or a programmer) knowledge of the database. On the other hand, automatically gathered statistics show an exact image of the database status. Thus, by computing inter-objects access frequencies for each structural relationship and each object at run-time, a better performance should be achieved.

# REFERENCES

[ANDE90], T.L. Anderson, A.J. Berre, M. Mallison, H.H. Porter III, B. Scheider, *The HyperModel Benchmark*, International Conference on Extending Database Technology, Venice, Italy, March 1990, pp. 317-331

[ANDR91a], T. Andrews, *Programming with Vbase*, In "Object-Oriented Databases with Applications to CASE, Networks and VLSI CAD", Edited by R. Gupta and E. Horowitz, Prentice Hall Series in Data and Knowledge Base Systems, 1991, pp. 130-177

[ANDR91b], T. Andrews, C. Harris, K. Sinkel, *ONTOS: A Persistent Database for C++*, In "Object-Oriented Databases with Applications to CASE, Networks and VLSI CAD", Edited by R. Gupta and E. Horowitz, Prentice Hall Series in Data and Knowledge Base Systems, 1991, pp. 387-406

[BANE87], J. Banerjee, H.-T. Chou, J.F. Garza, W. Kim, D. Woelk, N. Ballou, H.-J. Kim, *Data Model Issues for Object-Oriented Applications*, ACM Transaction on Office Information Systems, Vol. 5, No. 1, January 1987, pp. 3-26

[BATO85], D.S. Batory, W. Kim, *Modeling Concepts for VLSI CAD Objects*, ACM Transactions on Database Systems, Vol. 10, No. 3, September 1985, pp. 322-346

[BENZ90], V. Benzaken, C. Delobel, *Enhancing Performance in a Persistent Object Store: Clustering Strategies in $O_2$*, $4^{th}$ International Workshop on Persistent Object Systems, September 1990, pp. 403-412

[BERR91], A.J. Berre, T.L. Anderson, *The HyperModel Benchmark for Evaluating Object-Oriented Databases*, In "Object-Oriented Databases with Applications to CASE, Networks and VLSI CAD", Edited by R. Gupta and E. Horowitz, Prentice Hall Series in Data and Knowledge Base Systems, 1991, pp. 75-91

[CATT91], R.G.G. Cattell, *Object Data Management: Object-Oriented and Extended Relational Database Systems*, Addison-Wesley Publishing Company, 1991

[CHAB93], S. Chabridon, J.-C. Liao, Y. Ma, L. Gruenwald, *Clustering Techniques for Object-Oriented Database Systems*, 38th IEEE Computer Society International Conference, February 1993, San Francisco, pp. 232-242

[CHAN89a], E.E. Chang, *Effective Clustering and Buffering in an Object-Oriented DBMS*, University of California, Berkeley, Computer Science Division (EECS), Technical Report No. UCB/CSD 89/515, June 1989

[CHAN89b], E.E. Chang, R.H. Katz, *Exploiting Inheritance and Structure Semantics for Effective Clustering and Buffering in an Object-Oriented DBMS*, ACM SIGMOD International Conference on Management of Data, Portland, Oregon, June 1989, pp. 348-357

[CHAN90], E.E. Chang, R.H. Katz, *Inheritance in computer-aided design databases: semantics and implementation issues*, CAD, Vol. 22, No. 8, October 1990, pp. 489-499

[CHEN91], J.R. Cheng, A.R. Hurson, *Effective clustering of complex objects in object-oriented databases*, ACM SIGMOD International Conference on Management of Data, Denver, Colorado, May 1991, pp. 22-31

[DEUX90], O. DEUX et al., *The Story of $O_2$*, IEEE Transactions on Knowledge and Data Engineering, Vol. 2, No. 1, March 1990, pp. 91-108

[FORD88], S. Ford, J. Joseph, D.E. Langworthy, D.F. Lively, G. Pathak, E.R. Perez, R.W. Peterson, D.M. Sparacin, S.M. Thatte, D.L. Wells, S. Agarwala, *ZEITGEIST: Database Support for Object-Oriented Programming*, 2nd International Workshop on Object-Oriented Database Systems, Bad Münster am Stein-Ebernburg, FRG, September 1988, pp. 23-42

[GRUE91], L. Gruenwald, M.H. Eich, *MMDB Reload Algorithms*, ACM SIGMOD International Conference on Management of Data, Denver, Colorado, May 1991, pp. 397-405

[HE93], M. He, A.R. Hurson, L.L. Miller, D. Sheth, *An Efficient Storage Protocol for Distributed Object-Oriented Databases*, IEEE Parallel & Distributed Processing, 1993, pp. 606-610

[HUDS89], S.E. Hudson, R. King, *Cactis: A Self-Adaptive Concurrent Implementation of an Object-Oriented Database Management System*, ACM Transactions on Database Systems, Vol. 14, No. 3, September 1989, pp. 291-321

[HURS93], A.R. Hurson, S.H. Pakzad, J.-b. Cheng, *Object-Oriented Database Management Systems: Evolution and Performance Issues*, IEEE Computer, February 1993, pp. 48-60

[KATZ91], R.H. Katz, E.E. Chang, *Inheritance Issues in Computer-Aided Design Databases*, In "On Object-Oriented Database Systems", Edited by K.R. Dittrich, U. Doyal and A.P. Buchman, Springer-Verlag Topics in Information Systems, 1991, pp. 45-52

[KIM90a], W. Kim, J.F. Garza, N. Ballou, D. Woelk, *Architecture of the ORION Next-Generation Database System*, IEEE Transactions on Knowledge and Data Engineering, Vol. 2, No. 1, March 1990, pp. 109-124

[KIM90b], W. Kim, *Object-Oriented Databases: Definition and Research Directions*, IEEE Transactions on Knowledge and Data Engineering, Vol. 2, No. 3, September 1990, pp. 327-341

[MAIER86], D. Maier, J. Stein, A. Otis, A. Purdy, *Development of an Object-Oriented DBMS*, ACM OOPSLA '86 Proceedings, September 1986, pp. 472-482

[SRIN91], V. Srinivasan, M.J. Carey, *Performance of B-Tree Concurrency Control Algorithms*, ACM SIGMOD International Conference on Management of Data, Denver, Colorado, May 1991, pp. 416-425

[TSAN92a], M.M. Tsangaris, J.F. Naughton, *On the Performance of Object Clustering Techniques*, ACM SIGMOD International Conference on Management of Data, San Diego, California, June 1992, pp. 144-153

[TSAN92b], M.M. Tsangaris, *Principles for Static Clustering for Object Oriented Databases*, University of Wisconsin-Madison, Computer Science Department, Technical Report #1104, August 1992

[WILK88], W. Wilkes, *Instance Inheritance Mechanisms for Object Oriented Databases*, 2nd International Workshop on Object-Oriented Database Systems, Bad Münster am Stein-Ebernburg, FRG, September 1988, pp. 274-279